\documentclass{mn2e}
 \usepackage[dvips]{epsfig}
 \usepackage{amsfonts,amsbsy,amssymb}
\title [The optical counterpart of XTE J0929--314] 
{The optical counterpart of XTE J0929--314, the third transient 
   millisecond X-ray pulsar}
\author [A. B. Giles {\it et al.}]       
{A. B. Giles$^{1,2}$, J. G. Greenhill$^{1}$, K. M. Hill$^{1}$ and 
        E. Sanders$^{3}$\\
      $^{1}$ School of Mathematics and Physics, University of Tasmania, 
             GPO Box 252-21, Hobart, Tasmania 7001, Australia \\
      $^{2}$ Spurion Technology Pty. Ltd., 200 Mt. Rumney Road, Mt. Rumney, 
             Tasmania 7170, Australia \\
      $^{3}$ University of Tasmania Visitor } 

\date{Accepted 2005 June 6.
       Received 2005 May 3}
\pagerange{\pageref{firstpage}--\pageref{lastpage}}
\pubyear{2005}

\begin{document}
\maketitle

\label{firstpage}

\begin{abstract}
A blue and variable optical counterpart of the X-ray transient XTE J0929--314 
was identified on 2002 May 1. We conducted frequent {\it BVRI\/} broadband 
photometry on this object using the Mt Canopus 1-m telescope during May 
and June until it had faded to below 21'st magnitude. Nearly continuous 
{\it I\/} band CCD photometry on 2002 May 2, 3 \& 4 revealed a $\sim$ 10 per 
cent sinusoidal modulation at the binary period lasting $\sim$ 6 cycles during 
the latter half of May 2. The phase indicates that the modulation may be due 
to a combination of emission by a hot spot on the disc and X-ray heating of 
the secondary. The emission generally trended bluer with {\it B-I\/} 
decreasing by 0.6 magnitudes during the observations but there were anomalous 
changes in colour during the first few days after optical identification when 
the {\it I\/} band flux {\it decreased} slightly while fluxes in other bands 
{\it increased}. Spectral  analysis of the {\it BVRI\/} broadband photometry 
show evidence of a variable excess in the {\it R\/} \& {\it I\/} bands. 
We suggest that this may be due to synchrotron emission in matter flowing 
out of the system and note that similar processes may have been responsible 
for anomalous {\it V\/} \& {\it I\/} band measurements in 1998 of the persistent 
millisecond X-ray pulsar SAX J1808.4--3658.

\end{abstract}

\section{Introduction} 
X-ray heating of three regions is generally believed to contribute to optical 
variability in low mass X-ray binary (LMXB) systems. These are the accretion 
disc, a bright spot on the outer edge of the accretion disc due to 
inflowing material and the hemisphere of the companion facing the neutron 
star. In most LMXBs the reprocessed X-ray optical flux dominates the optical 
light from the rest of the system (van Paradijs 1983, van Paradijs \& 
McClintock 1995), particularly in the outburst phase. The companion itself 
may only be evident at a very faint level when the system is in quiescence. 
More recently it has become apparent that synchrotron emission from matter 
flowing out of the system via bipolar jets makes a highly variable 
contribution to radio and IR emission from many different classes of X-ray 
binaries (Fender 2003). In some cases this emission may extend into the 
optical region (Hynes et al. 2000). At least one other persistent millisecond 
X-ray pulsar, SAX J1808.4--3658, is known to have a transient IR excess and 
radio emission probably due to synchrotron processes (Wang et al. 2001).

On 2002 April 30 an X-ray transient was discovered by Remillard et al. (2002) 
using the All Sky Monitor (ASM) (Levine et al. 1996) on The Rossi X-ray Timing 
Explorer {\it RXTE} satellite. XTE J0929--314, the subject of this paper, was 
subsequently found to also be a millisecond X-ray pulsar by Remillard, Swank 
\& Strohmayer (2002). Using additional {\it RXTE} Proportional Counter Array 
(PCA) observations Galloway et al. (2002a; 2002b) reported a neutron star 
spin frequency of 185 Hz, a binary period of 2615-s and an implied companion 
mass of $\sim 0.008 M_{\odot}$, about 8.5 Jupiter masses. A blue and variable 
optical couterpart was suggested by Greenhill, Giles \& Hill (2002). 
This identification was supported by spectra obtained by Castro-Tirado et al. 
(2002) who found a number of emission lines superimposed on a blue continuum 
which is typical of soft X-ray transients in outburst. A coincident radio 
source was also reported by Rupen, Dhawan \& Mioduszewski (2002). 

XTE J0929--314 was the third transient millisecond X-ray pulsar to be 
discovered. The first (SAX J1808.4--3658) has been studied  extensively at 
all wavelengths (Giles, Hill \& Greenhill 1999;  Wang et al. 2001; Homer 
et al. 2002; Wachter et al. 2000; in 't Zand et al. 1998; Wijnands et al. 
2001; Markwardt, Miller \& Wijnands 2002; Wijnands \& van der Klis 1998; 
Chakrabarty \& Morgan 1998; Chakrabarty et al. 2003; Wijnands et al. 2003). 
The {\it V\/} band flux from SAX J1808.4--3658 decayed from $\sim 16.75-18.5$ 
mag in 10-d suggesting an e-folding time of 5-6 d.

Of the other four known millisecond X-ray pulsars only two have identified 
visible counterparts. XTE J1814--338 (Strohmeyer et al. 2003) has been 
identified by Krauss et al. (2003) but no detailed optical or IR 
observations are available. The more recent IGR J00291+5934 (Galloway et al. 
2005) has a detailed optical light curve (Bikmaev et al. 2005). 
The {\it R\/} band flux from this object decayed from $\sim 17.4-22.4$ mag 
in 30-d giving an e-folding time of $5.66\pm0.2$ d. Study of all these 
systems is expected to provide important information on the evolutionary 
path by which a conventional LMXB system might turn into a millisecond 
radio pulsar.  

In this paper we describe the optical variability of XTE J0929--314. The 
measurements were made in the {\it BVRI\/} bands during a period of 
$\sim$ 9 weeks following its discovery.

\section{Observations}
All the observations described in this paper were made using the 1-m 
telescope at the University of Tasmania Mt. Canopus Observatory. The CCD 
camera, its operating software (CICADA), the image reduction and analysis 
tools (MIDAS and DoPHOT) were identical to that described in Giles et al. 
(1999). The CCD camera contains an SITe chip which is a thinned back 
illuminated device providing 512 x 512 pixels with an image scale of 
$0.434 \arcsec $ pixel$^{-1}$. Cousins standard {\it BVRI\/} filters 
(Bessell 1990) were used for the observations. The data were calibrated 
using a sequence of observations of 11 standard stars within the RU149D 
and PG1047 fields of Landolt (1992) to derive the magnitudes of five local 
secondary standards close to XTE J0929--314 and within the CCD frame. 
These local standards are marked as stars 1-5 on the finder chart in 
Fig. 1 and we tabulate their derived magnitudes in Table 1. The magnitudes 
for XTE J0929--314 were then obtained using differential photometry 
relative to these local secondary standards. We did not use the same 
stars for all colours but used a combination of three of the five as 
indicated in Table 1. This multi-star process revealed that differential 
colour corrections are negligible. A few observations were interrupted 
or terminated early by the arrival of clouds. The complete data set 
from 2002 May 1 to July 1 [HJD 2452(396) - 2452(457)] is detailed in 
Tables 2 \& 3 and forms the subject of this paper.

\begin{figure}
\epsfig{file=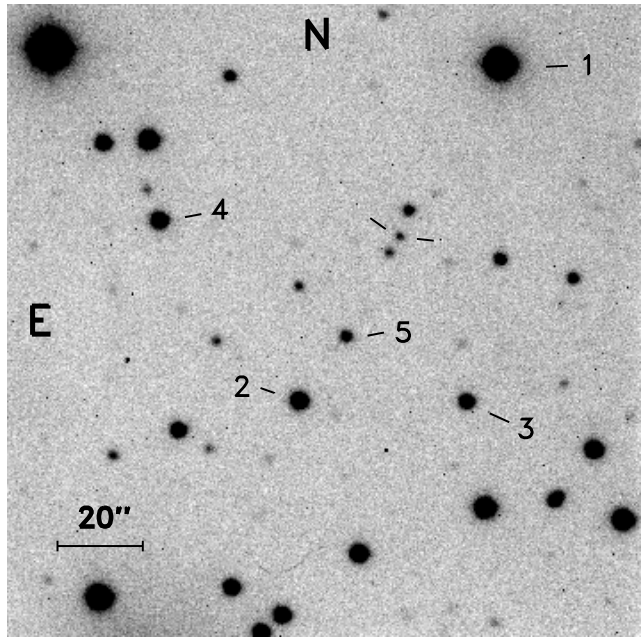, width=8.4cm } 
  \caption{A finder chart for XTE J0929--314. This is an {\it I\/} band 
  CCD image from May 2 (HJD 2452396.84424) when the source was at an 
  {\it I\/} magnitude of 18.75. The five local secondary standards listed in 
  Table 1 are marked with the numbers 1-5. We note that there is a very 
  bright star with {\it B\/} $\sim$ 12.6 just off the image to the West 
  of XTE J0929--314. }
\end{figure}

\begin{table}
 \caption{The magnitudes of local standard stars 1-5 in Fig. 1.}
 \label{symbols}
 \begin{tabular}{@{}ccccc}
 \hline
     Star No.      &  {\it B\/}  &  {\it V\/}  &  {\it R\/}  &  {\it I\/}  \\
 \hline
        1          &   15.029    &   14.297    &   13.878    &   13.483    \\
        2          &   17.583    &   16.838    &   16.403    &   16.003    \\
        3          &   18.542    &   17.636    &   17.149    &   16.714    \\
        4          &   18.083    &   17.106    &   16.567    &   16.154    \\
        5          &   19.443    &   18.601    &   18.194    &   17.781    \\
%       6          &   17.645    &   16.621    &   16.062    &   15.487    \\
%    Mean          &  16.084(30) &  17.145(35) &  17.022(30) &  16.610(35) \\
  Stars used       &   1, 2, 3   &   2, 3, 4   &   2, 3, 5   &   2, 3, 5   \\    
  Composite error  &    0.030    &    0.035    &    0.030    &    0.035    \\
 \hline
 \end{tabular}
 \medskip
\end{table}

\begin{table*}
\centering
\begin{minipage}{175mm}
\begin{center}
 \bf Table 2. \rm A journal of the {\it B\/}, {\it V\/} \& {\it R\/} band  observations. 
 \vspace*{0.125cm}
 \label{symbols}
 \begin{tabular}{@{}llrllrllr}
 \hline
 $HJD^{a}$  &                & Int. & $HJD^{a}$  &                & Int.  & $HJD^{a}$  &                & Int. \\
 (-2452000) & {\it B\/} mag. & (s)  & (-2452000) & {\it V\/} mag. & (s)   & (-2452000) & {\it R\/} mag. & (s)\\
 \hline
  395.92676  &  19.18(4)  &  300  &  396.03104  &  19.32(3)   &  600  &  398.98608  &  19.07(4)  &  600  \\
  395.92676  &  19.18(4)  &  300  &  399.00464  &  19.04(3)   &  600  &  405.02222  &  18.94(4)  & 1200  \\
  395.94162  &  19.17(4)  &  120  &  404.96179  &  18.71(2)   & 1200  &  405.88013  &  18.83(2)  & 1200  \\
  395.99146  &  19.07(9)  &  120  &  405.86371  &  18.68(1)   &  900  &  405.98022  &  18.87(2)  & 1200  \\
  395.99707  &  19.05(8)  &  120  &  405.89260  &  18.67(1)   &  900  &  407.94819  &  18.93(3)  &  600  \\
  396.02051  &  19.15(6)  &  300  &  405.99594  &  18.70(2)   &  900  &  407.95190  &  18.93(3)  &  600  \\
  398.99536  &  18.88(3)  &  600  &  406.84912  &  18.63(5)   &  600  &  409.04370  &  19.11(4)  &  600  \\
  405.90308  &  18.45(1)  &  900  &  406.85823  &  18.63(3)   &  600  &  409.06494  &  19.05(4)  &  600  \\
  406.00757  &  18.50(3)  &  900  &  406.86554  &  18.69(3)   &  600  &  409.87354  &  19.16(3)  &  900  \\
  407.96997  &  18.50(2)  &  600  &  406.87273  &  18.66(4)   &  600  &  413.85767  &  19.07(4)  &  729  \\
  409.03638  &  18.58(2)  &  600  &  406.88182  &  18.64(4)   &  900  &  425.97276  &  19.82(9)  & 1200  \\
  413.88916  &  18.76(2)  &  900  &  406.89266  &  18.69(3)   &  900  &  426.87783  &  19.73(6)  & 1200  \\
  426.91064  &  19.47(4)  & 1200  &  406.94484  &  18.67(2)   &  900  &  431.92334  &  19.82(5)  & 1200  \\
  431.93872  &  19.60(4)  & 1200  &  406.95277  &  18.71(2)   &  424  &  456.88272  &  22.21(86) &  300  \\
  456.86987  &  21.38(41) &  300  &  406.96392  &  18.67(2)   &  900  &             &            &       \\
  456.91085  & 22.19(119) &  300  &  407.94437  &  18.74(2)   &  600  &             &            &       \\
             &            &       &  409.02885  &  18.92(2)   &  600  &             &            &       \\
             &            &       &  409.97841  &  18.87(6)   &  600  &             &            &       \\
             &            &       &  413.00105  &  18.80(6)   &  900  &             &            &       \\
             &            &       &  413.86740  &  19.00(6)   &  446  &             &            &       \\
             &            &       &  413.87759  &  18.97(3)   &  900  &             &            &       \\
             &            &       &  426.89536  &  19.62(5)   & 1200  &             &            &       \\
             &            &       &  431.90887  &  19.91(5)   & 1200  &             &            &       \\
             &            &       &  456.87879  &  22.89(137) &  300  &             &            &       \\
             &            &       &  456.91490  &  22.19(121) &  300  &             &            &       \\
 \hline
 \end{tabular}
\end{center}
\hspace*{2.25cm} {\em a} Times of mid integration. Some integrations terminated early.
\end{minipage}
\end{table*}

\begin{table}
% \caption{A journal of the {\it I\/} band observations.}
%\begin{left}
%\bf Table \ 3. \rm \ A \ journal \ of \ the \ {\it I\/} \ band \ observations.
\bf Table 3. \rm A journal of the {\it I\/} band observations. \ \ \ \ \ \ \ \ \ \ \ \ \ \ \ \ \ 
% \end{left}
 \label{symbols}
 \begin{tabular}{@{}llrllr}
 \hline
 $HJD^{a}$  &                & Int. & $HJD^{a}$  &                & Int. \\
 (-2452000) & {\it I\/} mag. & (s)  & (-2452000) & {\it I\/} mag. & (s)  \\
 \hline
  396.08301  &  18.78(7)  &  600  &  408.85669  &  18.34(3)  &  600  \\
  396.09229  &  18.66(7)  &  600  &  408.87915  &  18.34(3)  &  600  \\
  396.09961  &  18.77(8)  &  600  &  408.89258  &  18.35(2)  &  600  \\
  396.10718  &  18.86(10) &  600  &  409.02100  &  18.36(3)  &  600  \\
  396.97188  & $18.81^{b}$ &      &  409.04932  &  18.40(20) &  300  \\
  397.99811  & $18.90^{b}$ &      &  409.05762  &  18.43(5)  &  600  \\
  398.96029  & $18.76^{b}$ &      &  409.86133  &  18.54(4)  &  600  \\
  404.90161  &  18.22(3)  &  900  &  412.88794  &  18.52(6)  &  900  \\
  404.94751  &  18.18(3)  &  900  &  412.90015  &  18.44(5)  & 1200  \\
  404.97412  &  18.20(3)  &  900  &  412.93945  &  18.38(4)  & 1200  \\
  404.98730  &  18.23(3)  &  900  &  412.95410  &  18.38(3)  &  903  \\
  404.99805  &  18.20(3)  &  900  &  412.96509  &  18.35(3)  &  900  \\
  405.00879  &  18.26(3)  &  900  &  412.97876  &  18.39(3)  &  900  \\
  405.91406  &  18.17(2)  &  600  &  412.98999  &  18.32(4)  &  900  \\
  405.92236  &  18.09(3)  &  600  &  413.84814  &  18.54(3)  &  900  \\
  405.93091  &  18.07(3)  &  600  &  413.90015  &  18.54(2)  &  900  \\
  405.93921  &  18.16(2)  &  600  &  413.93921  &  18.53(3)  &  900  \\
  405.94678  &  18.17(2)  &  600  &  413.97998  &  18.51(3)  &  900  \\
  405.95386  &  18.18(2)  &  600  &  414.02148  &  18.47(3)  &  900  \\
  405.96143  &  18.24(3)  &  600  &  414.06055  &  18.53(4)  &  900  \\
  405.96899  &  18.18(3)  &  600  &  425.95044  &  19.35(12) &  600  \\
  406.01758  &  18.29(7)  &  600  &  425.96143  &  19.66(13) &  600  \\
  407.06348  &  18.21(4)  &  900  &  425.98877  &  19.61(11) & 1200  \\
  407.07397  &  18.18(6)  &  900  &  426.86182  &  19.69(9)  & 1200  \\
  407.08496  &  18.16(4)  &  900  &  426.98071  &  19.53(9)  & 1200  \\
  407.92944  &  18.18(3)  &  600  &  431.89160  &  19.56(6)  & 1200  \\
  407.97754  &  18.32(4)  &  600  &  431.96948  &  19.78(10) & 1200  \\
  408.02051  &  18.21(4)  &  600  &  431.98413  &  19.59(11) & 1200  \\
  408.06104  &  18.14(4)  &  600  &  456.87491  &  20.79(45) &  300  \\
  408.07007  &  18.18(4)  &  900  &  456.90527  &  20.84(63) &  300  \\
 \hline
 \end{tabular}
 \medskip
{\em a} Times of mid integration. Some integrations terminated early. \\
{\em b} Average magnitude and mid time for entire nights data. 
\end{table}

\section{Results}

\subsection{Source position}
To determine the source position we selected the nearest 9 stars to the
candidate from the Hubble Guide Star Catalogue 2.2 downloaded 
from the NASA HEASARC web site. We used the CCD coordinates of these stars 
and the proposed candidate on the best quality {\it I\/} band image from 
May 2 (HJD 397) to derive an accurate source position for XTE J0929--314 
of R.A. 9h 29m 20$\fs19$  Dec. $-31\degr 23\arcmin 3\farcs2$ with a 
relative error of $\pm 0\farcs1$ (equinox J2000.0). This is 0$\farcs7$ 
from our initial position estimate in Greenhill et al. 2002 and only 
$\sim$0$\farcs2$ away from the radio position given by Rupen et al. (2002). 
{\it Chandra} observations confirmed that the X-ray source was 
$\sim$ 1$\farcs25$ from our optical position (Juett, Galloway \& 
Chakrabarty 2003). We note that this difference is twice their quoted error. 

In Fig. 1 we provide a finder chart for XTE J0929--314 constructed from 
an {\it I\/} band CCD image from May 2. The object is the central star in 
a line of three and for the first few weeks was seen to be of similar 
brightness to its two neighbours.

\subsection{The X-ray light curve}
XTE J0929--314 is one of the faintest transients to be found by the ASM 
experiment on {\it RXTE} and this detection was only possible due to its 
angular distance from the Galactic Centre region which minimises bright 
source confusion and consequent positional uncertainties. Our optical 
candidate was only $0\farcm5$ from the X-ray position of Remillard et al. 
(2002). The transient was discovered near the time of peak X-ray flux, 
towards the end of April. It was then observed periodically with the PCA 
experiment on {\it RXTE} [obs id 70096-03-(**-**)] as part of a proprietary 
TOO campaign (Galloway et al. 2002b; Juett et al 2003). In the top panel 
of Fig. 2 we show the ASM intensity history for this transient up to just 
past the time when regular PCA observations commenced. Note that there is 
a single early PCA observation at HJD 397.0. The ASM points are plotted 
as daily averages since the data for individual dwell cycles are too noisy 
for such a relatively faint source. The 38 PCA values are averages over 
individual PCA observation id's which typically last 1000-4000 seconds. 
The two X-ray light curves in Fig 2. are similar to fig. 1 in Galloway 
et al. (2002b) except that both vertical axes are now drawn with 
'X-ray magnitude' scales. The final three PCA observations are not plotted 
since only upper limit detections exist after HJD 445.0.

\begin{figure}
\epsfig{file=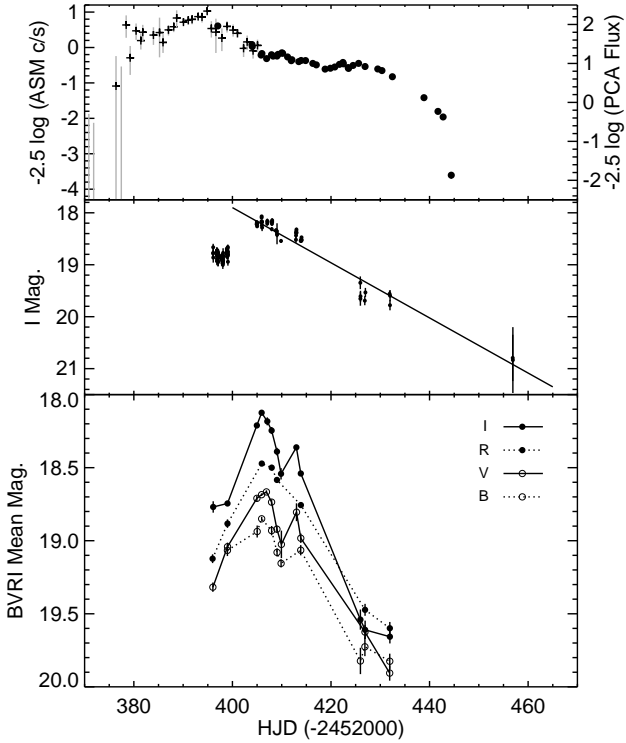, width=8.4cm }
  \caption{The {\it RXTE} ASM (+) and PCA ($\bullet$) light curves for 
  XTE J0929--314 are shown in the upper panel from April 5 to June 19 
  (HJD 370 - 445). The right hand PCA axis is in units of 
  10$^{-10}$ erg cm$^{-2}$ s$^{-1}$ for the 2-10 keV band. 
  The centre panel light curve shows all the {\it I\/} data with 
  a linear fit to the decay interval. The lower panel shows the 
  {\it BVRI\/} band light curves for the average flux on each night. } 
\end{figure}

\subsection{The optical light curves}
In Fig. 2 we plot our sets of {\it BVRI\/} band data from 
Tables 2 \& 3. The tight clusters of {\it I\/} data on May 2, 3 \& 4 
(HJD 397, 398 \& 399) are shown in greater detail in Fig. 4. There is 
evidence of variability in the range 0.05 to 0.1 magnitudes on a 
timescale of hours during several nights when long runs of {\it I\/}
band measurements were made. Greenhill et al (2002) reported variability 
of up to $\sim 0.5$ magnitudes on the first night (HJD 396) of our 
observations. Preliminary analysis suggested that a short duration 
(timescale $\sim 30$ minutes) $\sim 0.5$ magnitude flare occurred in 
{\it V \/} on this night. Subsequently we became aware of water vapour 
condensation on the filter during this "event" and we are now doubtful of 
the reality of the flaring. In the following section we describe evidence for 
a low amplitude orbital period modulation seen in the {\it I\/} band flux
at about HJD 397. It is of interest to note that no significant variations were 
seen in X-rays during the {\it RXTE} PCA observations (Galloway et al. 2002b). 
The upper limit to the orbital period modulation of the 2-10 keV X-ray 
flux was $<1.1$ per cent (3$\sigma$) (Juett et al. 2003). 

In Fig. 3 we plot the mean {\it B-V\/} and {\it V-I\/} colour indices for 
the nights when these three colours were measured. In order to minimise the 
effects of source variability the colour indices for each night are derived 
from one or more measurements in the different colours taken within a time 
interval of $\sim$ 1 hour. The overall trend was for the spectrum to become 
hotter (more blue) over the principal five weeks of observation. 
This suggests, assuming that most of the light comes from an accretion 
disc, a trend towards increasing disc temperature as the accreting matter 
diffuses inwards. There was also a brief decrease in {\it B-V\/} \& {\it V-I\/} 
(increase in colour temperature) between 2002 May 1 and May 4 (HJD 396 - 399), 
followed by a recovery to the overall trend line. This can be seen in 
Fig. 2 where, for the first few days following optical identification, 
the {\it I\/} band flux {\it decreased} slightly while fluxes in the other 
bands {\it increased}.

\begin{figure}
\epsfig{file=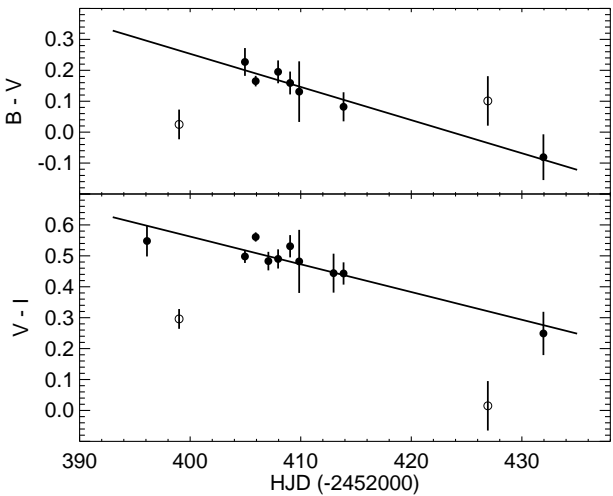, width=8.4 cm}
  \caption{Time dependence of the colour indices {\it B - V} (top panel) 
  and {\it V - I} (bottom panel).Both data sets are shown with a linear fit. 
  The anomalous points on HJD 399 and 427 (marked with $\circ$ symbols) 
  have been excluded from the fitting process.}
\end{figure}

The light curves in {\it BVRI\/} are approximately triangular in profile, on 
a linear scale, and similar to, but delayed by, $\sim$ 13 days relative to the 
X-ray light curve. During the time interval May 1 to May 11 (HJD 396 - 406) 
the {\it BVRI\/} fluxes {\it increased} by $\sim 75$ per cent while 
the X-ray flux {\it decreased} by $\sim 50$ per cent. We can think of no 
physical process whereby X-ray emission can \it lead \rm optical emission 
by 13 days. In SAX J1808.4--3658, the optical decline \it preceded \rm 
the X-ray by $3\pm 1d$ (Giles et al. 1999). Since we have no information 
on the optical flux during April we conclude that the apparent similarity 
between the optical and X-ray light curves is coincidental. The optical 
decay evident in the centre panel of Fig. 2 has an e-folding time of 
$22.2\pm1.1$ d. This appears to be qualitatively different to that for 
SAX J1808.4--3658 and IGR J00291+5934

\subsection{Search for binary modulation}

On the nights of May 2, 3 \& 4 (HJD 397, 398 \& 399) we monitored the 
object at {\it I\/} continuously in an attempt to observe an orbital 
period modulation. These data are plotted in Fig. 4. In mid-May Galloway 
et al. (2002a) detected a 2614.75(15) s orbital period modulation of the 
X-ray pulsation frequency. No amplitude modulation of the X-ray flux 
was detected. Their orbit ephemeris placed the neutron star 
on the far side of the companion at 2002 May 11.4941(2) UT (HJD 405.9941).
The solid sine curves in the lower part of each panel in Fig. 4 represent 
the modulation (arbitrary amplitude) expected from X-ray heating of the 
companion star as in SAX J1808.4--3658 (see fig. 1 in Giles et al. 1999 
and fig. 2 in Homer et al. 2002). In Fig. 4 we also show the light 
curves for a nearby 'constant' comparison star ($\sim6\arcsec$ to the NNW 
in Fig. 1, {\it I\/} mag 17.11) which has been shifted by an arbitary amount. 
This star does not show the modulation evident for XTE J0929--314 
in the top panel. All three observing runs terminated close to the 
telescope elevation limit but the source exhibits significant variability which 
is not apparent for the plotted comparison star or for others not 
shown here.

\begin{figure}
\epsfig{file=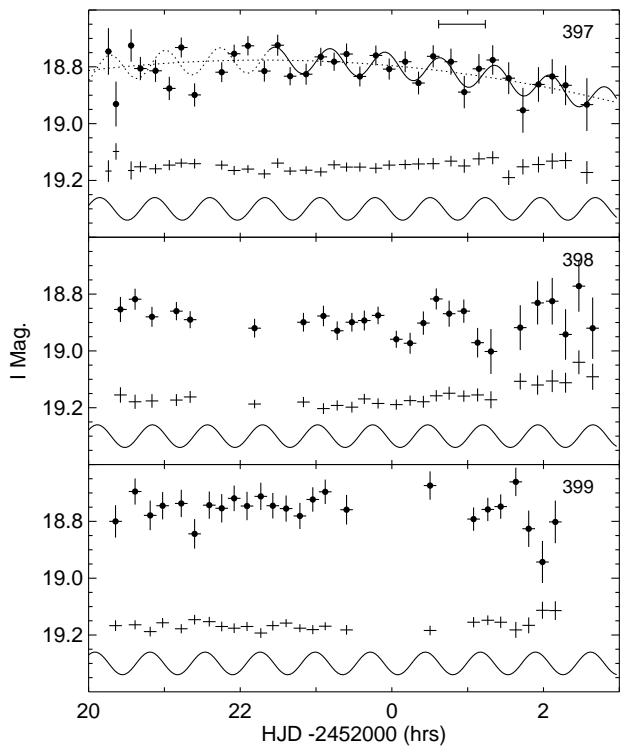, width=8.4cm}
  \caption{The {\it I\/} band light curves for May 2, 3 \& 4 (HJD 397, 398 
  \& 399) for XTE J0929--314 ($\bullet$) and a nearby comparison star (+). 
  All integrations are 600-s. The number to the right of each panel gives 
  the truncated HJD starting at zero hours. The sine curve marks the possible 
  optical modulation which might be expected to occur in anti-phase to the 
  X-ray ephemeris. The sine modulation parameters for the fit on May 2 
  (HJD 397) are given in the main text and the dotted line represents a 
  second order polynomial fit to de-trend the data. A horizonal line in the 
  top panel marks the duration of the first PCA X-ray observation. }
\end{figure}

The light curve for the last four hours of the night of May 2 (HJD 397) shows 
clear indications of modulation at or near the orbital period together 
with slow changes in the average magnitude. We corrected the data commencing 
from HJD 396.92 ($\sim 22$h UT) for these slow changes using a second order 
polynomial and used the Q method (Warner \& Robinson, 1972) to search the 
de-trended data for periodicity in the range 0.01 to 0.05 days. There is a 
strong single peak corresponding to a period of $0.030 \pm 0.001$ days. 
This is consistent with the orbital period of 0.030263 days discovered by 
Galloway et al. (2002b).

\begin{figure}
\epsfig{file=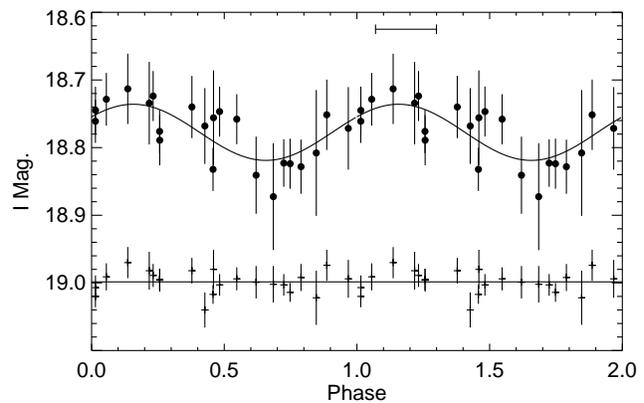, width=8.4cm}
  \caption{The {\it I\/} band light curve for part of the night of 2002 May 2 
  (HJD 397) folded at the orbital period. The solid curve is a best fit 
  sinusoid at the X-ray period. The lower set of points show the scatter 
  for the constant nearby star. The horizontal bar represents the 600-s 
  duration of the individual integrations.}
\end{figure}

In Fig. 5. we plot the corrected light curve folded at the X-ray period and 
ephemeris where phase zero is defined as the time when the companion is at 
its greatest distance from the observer i.e. it lies beyond the neutron star. 
The vertical error bars are those generated by the DoPHOT photometry which 
appear to be slightly over-estimated. To clarify this we repeated the phase 
folding process for the nearby comparison star, using the same X-ray 
ephemeris, so that the real scatter for a 'constant' source can be seen. 
These points are plotted in the lower part of Fig. 5 and have a  slightly 
different offset to that used in Fig. 4.  The formal DoPHOT average error 
for these 22 points is 0.021 mag but they have a $\pm 1 \sigma$ scatter of 
only 0.017 mag so the XTE J0929--314 error bars in Figs. 4 \& 5 are probably 
$\sim$ 20 per cent too large. 

The modulation is approximately sinusoidal with amplitude $\sim 0.09$ 
magnitudes peak to peak and maximum at phase $0.19 \pm 0.05$. Hence the 
modulation is unlikely to be due to X-ray heating of the companion as this 
would have maximum light at phase zero. In this respect XTE J0929--314 differs 
from SAX J1808.4--3658 in which the orbital period optical modulation had a 
maximum at phase zero (Giles et al., 1999). Nor is it likely that the 
modulation is due to emission from a hot spot on the disc since this would 
have a maximum at a phase between $\sim 0.3 - 0.5$. Perhaps both processes 
contribute.
 
The first PCA X-ray observation (obs. 01-00) occurred during the time we 
detected an orbital period modulation on the night of 2002 May 2 (HJD 397) 
and its duration is shown in the top panel of Fig. 4. However, this 
observation consists of crossed slews to determine the X-ray source 
position and is thus not suitable for modulation analysis. As noted 
earlier, no X-ray amplitude modulation was reported in any of the many 
following {\it RXTE} PCA observations (Juett et al. 2003).

\subsection{Spectral changes}
There are many occasions within Tables 2 \& 3 where we have {\it BVRI\/} 
values on the same night but some caution is required in combining these 
into broadband spectra due to the variability detected on several nights. 
In most instances the different colour averages for each night are derived 
from one or more measurements taken in a time interval of $\sim$ 1 hour. 
It should be clearly noted that the central wavelength and bandwidth for 
the {\it R\/} \& {\it I\/} filters differs between the Cousins and older 
Johnston systems and this can affect the apparent spectral shape. 
Here we use the Cousins {\it R\/} \& {\it I\/} parameters. In Fig. 6 we 
plot the broadband {\it BVRI\/} spectra from 8 nights, 2002 May 1, 4, 11, 
13, 14, 19 \& June 1, 7 (HJD 396, 399, 406, 408, 409, 414, 427 \& 432).

\begin{figure}
\epsfig{file=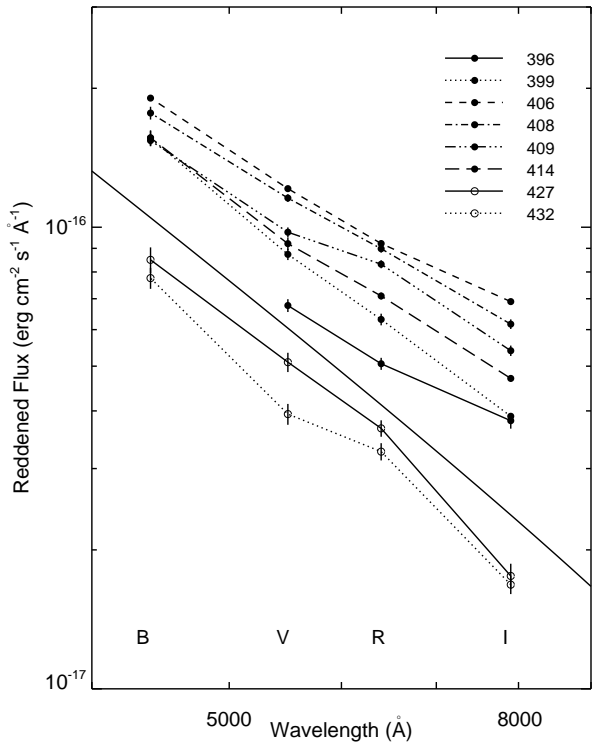, width=8.4cm }
  \caption{The {\it BVRI\/} broadband spectra for XTE J0929--314. Each day's 
  spectrum is identified by the start of the HJD falling within the daily 
  $\pm$ 4 h. observing window. The lines connect to the mean flux values 
  on each night and many error limits are smaller than the points representing 
  the measurements. The full width diagonal solid line represents a simple 
  power law disc emission model with exponent -3, arbitrary amplitude and 
  interstellar reddening corresponding to $A_{V}=0.42$. }
\end{figure}

Also shown is a curve representing a power law approximation to the 
emission from an optically thick, X-ray heated disc. The distribution is 
given by the equation $F_{\lambda} \propto \lambda^{-3}e^{-A_{\lambda } 
/1.086}$ where $F_{\lambda }$ is the reddened flux at wavelength $\lambda $ 
and $A_{\lambda }$ is the wavelength dependent reddening correction toward 
the source. The amplitude is arbitrary and the spectrum is reddened assuming 
interstellar extinction $A_{V}=0.42$. This value is scaled from the 
estimated value $A_{V}=0.68$ for  SAX J1808.4--3658 (Wang et al., 2001) using 
the integrated column densities, $N_H \approx 1.3 \times 10^{20} cm^{-2}$ 
for SAX J1808.4--3658 (Gilfanov et al., 1998) and $N_H  
\approx 7.6 \times 10^{20} cm^{-2}$ for XTE J0929--314 (Juett et al., 2003). 
A similar value for $A_{V}$ is obtained by using the relationship between 
$A_{V}$ and $N_H$ given by Predehl \& Schmitt (1995).

We have no information on the flux in {\it B\/} for the first night (HJD 396) 
but it is clear that the spectrum was heavily reddened on that occasion. 
On subsequent nights the spectra were generally steeper and had an approximately  
power law distribution. There was, however, a highly variable red excess 
above the power law. The excesses are not sensitive to the assumed value of 
$A_{V}$. On HJD 399 the excess was near zero (similar to the canonical disc 
power law distribution) and, as noted in section 3.3, the spectrum was 
anomalously blue. On HJD 409 (and possibly HJD 432 although with less 
statistical significance) the {\it R\/} band was strongly enhanced consistent 
with strong Balmer line ($H\alpha$) emission. Balmer emission cannot however 
account for the variable excesses in {\it I\/}. Measurements errors were 
relatively large for the last two nights in Fig. 6 (HJD 427 and 432) but it 
is clear that the red excess had disappeared and that the spectra were 
steeper (bluer) than during earlier observations.

\section{Discussion}
Several features distinguish XTE J0929--314 from SAX J1808.4--3658. 
Firstly, the maximum of the orbital period modulation occurs at phase 
$0.19 \pm 0.05$ rather than at phase zero. This points to an 
origin of X-ray heating in SAX J1808.4--3658 but the situation is more 
complex in XTE J0929--314. Galloway et al. (2002b) have shown that the 
companion in XTE J0929--314 is probably a very low mass ($\sim 0.008  
M_{\odot}$) helium white dwarf. The companion in SAX J1808.4--3658 is 
believed to be a $\sim 0.05M_{\odot}$ brown dwarf (Bildsten \& 
Chakrabarty, 2001). The Roche lobe radii of the companion stars will 
therefore differ by almost an order of magnitude substantially reducing 
the X-ray radiation reprocessed by the companion in XTE J0929--314. We 
have estimated the amplitude $L_{OA}$ of optical modulation due to heating 
of the companion star in XTE J0929--314 using the relation 
$L_{OA} = L_{XA}/L_{XB}(d_B/d_A)^{2}(R_A/R_B)^{2}L_{OB}$ where $L_{OB}$ 
is the optical modulation observed for SAX J1808.4--3658 (Giles et al, 
1999), $L_{XA}/L_{XB}$ is the ratio of the X-ray fluxes (Gilfanov et al., 
1998, Juett et al., 2003), $d_B/d_A$ is the ratio of distances between 
the neutron stars and their companions and $R_A/R_B$ is the ratio of the 
Roche Lobe radii for the two systems. The estimated orbital modulation 
is $\sim 25$ per cent of that observed suggesting that, if it is 
generated thermally, much of the emission comes from a hot spot on 
the disc. This provides an explanation for the observed maximum phase 
which lies mid-way between that expected from heating of the companion 
and from hot spot emission. 
 
Turning to the spectral characteristics the broadband spectra from 
SAX J1808.4--3658 are well fitted by smoothly varying functions derived 
from a thermal disc model (Wang et al., 2001). However, there are 
signifiant, time dependent, {\it R\/} \& {\it I\/}  band excesses in 
our XTE J0929--314 spectra. Nothing like this has been reported for 
SAX J1808.4--3658 although Wang et al. (2001) reported a strong near 
IR {\it JHK\/} excess on one occasion and this may well have extended 
to optical wavelengths. We have insufficient information to explain our 
observed spectral variability. As noted in section 3.5, variable Balmer 
emission may be responsible for the {\it R\/} band excesses but cannot 
contribute to the excesses in {\it I\/}. The remarkable changes in the 
red excesses between HJD 396 and 399 and between HJD 409 and 414 might 
perhaps be due to diffusion inwards of cool matter from a brief 
enhancement of mass transfer onto the disc from the companion star. 
Alternatively, the red excesses may be due to transient synchrotron 
emission from matter flowing out of the system via bipolar jets. 
The synchrotron spectrum is cut off at wavelengths shorter than {\it R\/}. 

As noted in section 1, many different classes of X-ray binaries emit 
synchrotron radiation (Fender 2003). Rupen et al. (2002) identified a 
weak ($\sim 0.35 \pm 0.07$ mJy) 4.86 GHz radio source at the SAX J0929-314 
position on 2002 May 3 and 7. Unfortunately there do not appear to have 
been any IR observations during this outburst. We hypothesise that a 
rather similar phenomenon may have occurred during the 1998 outburst 
of the accreting millisecond pulsar SAX J1808.4--3658. Wang et al (2001) 
noted that the {\it V\/} \& {\it I\/} band fluxes measured by the JKT 
1-m telescope on 1998 April 18.2 were about 0.2 magnitudes brighter than 
they were $\sim$ 0.5 days later when measured by the Mt Canopus 1-m 
telescope. They assumed that the discrepancy was due to calibration 
uncertainties in the JKT data. There was a clear IR ({\it JHK\/}) excess 
measured by the UKIRT telescope at 1998 April 18.6 just 0.4 days after 
the JKT measurements. Wang et al. (2001) proposed a synchrotron origin 
for this IR excess. We suggest that the synchrotron excess was also 
present at the time of the JKT measurements and that it extended into 
the optical bands. It had disappeared at the time of the Mt Canopus 
measurements a few hours later.

\section{Conclusions}
The optical counterpart of XTE J0929--314 was variable on all 
timescales down to a few hours during the 2002 May observations. 
On one occasion lasting $\sim 4$ hours the {\it I\/} band flux was modulated 
at the orbital period with amplitude $0.09 \pm 0.01$ magnitudes. 
No variability was apparent in the X-ray measurements (Galloway et al, 2002b). 
The peak of the orbital modulation occurs at a phase of $0.19 \pm 0.05$ 
relative to the X-ray ephemeris and appears to rule out X-ray heating of 
the companion as the source of the modulation unless it is combined with 
emission from a hot spot on the disc.

Broad band {\it BVRI\/} spectra taken on 8 nights have an approximately 
power law distribution as expected for an optically thick accretion 
disc but with variable excesses in {\it R\/} \& {\it I\/}. Overall these 
excesses declined and the spectra steepened (became bluer) during the period 
of the observations. While variable $H_\alpha$ emission may be responsible 
for some of the excess in {\it R\/}, another explanantion is required for 
the {\it I\/} band enhancements. We suggest they may be due to emission 
from cool matter in the outer part of the disc following a transient 
episode of mass transfer from the companion. Alternatively, variable 
synchrotron emission, cut off at {\it R\/} band wavelengths, contributes 
to the emission spectrum.

There is a clear need for fast follow-up optical, IR and radio observations 
of millisecond X-ray pulsars. These should include polarimetry and high 
time resolution optical and near IR photometry to test the synchrotron 
emission hypothesis. The feasibility of high speed photometry at the 
pulsar spin frequency might also be investigated. Facilities for immediate 
data reduction and generation of light curves are essential in order to 
optimise observing strategies for these highly variable objects.

\section{Acknowledgements}
We thank Ron Remillard for timely information on the occurrence of this 
transient and Duncan Galloway for providing PCA data. This research 
has made use of data obtained through the High Energy Astrophysics Science 
Archive Research Center Online Service, provided by the NASA / Goddard 
Space Flight Center. We thank Don Melrose, Mark Walker and the referee
for helpful comments and  gratefully acknowledge financial support for the 
Mt Canopus Observatory by Mr David Warren. ABG thanks the University of 
Tasmania Antarctic CRC for the use of computer facilities.

\label{lastpage}


\begin{thebibliography}{99}
\bibitem{b1}  Bessell M.S., 1990, PASP, 102, 1181
\bibitem{b2}  Bikmaev I. et al., 2005, ATEL 395
\bibitem{b3}  Bildsten L., Chakrabarty D., 2001, ApJ, 557, 292
\bibitem{b4}  Castro-Tirado A.J. et al., 2002, IAU Circ. 7895
\bibitem{b5}  Chakrabarty D., Morgan E.H., 1998, Nature, 394, 346 
\bibitem{b6}  Chakrabarty D., Morgan E.H., Muno M.P., Galloway D.K.,
                Wijnands R., van der Klis M., Markwardt C.B.,
                2003, Nature, 424, 42
\bibitem{b7}  Fender R.P., 2003, in Fender R.P., Macquart J-P.,
                eds, Circular Polarisation from Relativistic Jet Sources, 
                Ap\&SS, 288, 79
\bibitem{b8}  Galloway D.K., Morgan, E.H., Remillard R.A., Chakrabarty D.,
                2002a, IAU Circ. 7900
\bibitem{b9}  Galloway D.K., Chakrabarty D., Morgan, E.H., Remillard R.A., 
                2002b, ApJ, 576, L137
\bibitem{b10} Galloway D.K., Markwardt C.B., Morgan E.H., Chakrabarty D., 
                Strohmayer T.E., 2005, submitted ApJ,  (astro-ph/0501064)
\bibitem{b11} Giles A.B., Hill K.M., Greenhill J.G., 1999, MNRAS, 304, 47
\bibitem{b12} Gilfanov M., Revnivtsev M., Sunyaev R., Churazov E., 1998, 
                A\&A, 338, L83
\bibitem{b13} Greenhill J.G., Giles A.B., Hill K.M., 2002, IAU Circ. 7889
\bibitem{b14} Homer L., Charles P.A., Chakrabarty G., van Zyl L., 2002, 
                MNRAS, 325, 147
\bibitem{b15} Hynes R.I., Mauche C.W., Haswell C.A., Shrader C.A., Cui W., 
                Chaty S., 2000, ApJ, 539, L37
\bibitem{b16} in 't Zand J.J.M., Heise J., Muller J.M., Bazzano A., Cocchi M., 
                Natalucci I., Ubertini P., 1998, A\&A, 331, L25
\bibitem{b17} Juett A.M., Galloway D.K., Chakrabarty D., 2003, ApJ, 587, 754	
\bibitem{b18} Krauss M.I., Dullighan A., Chakrabarty D., van Kerkwijk M.H., 
                Markwardt C.B., 2003, IAU Circ. 8154
\bibitem{b19} Landolt A.U., 1992, ApJ, 104, 340
\bibitem{b20} Levine A.M., Bradt H., Cui W., Jernigan J.G., Morgan E.H., 
                Remillard R., Shirey R. E., Smith D.A., 1996, ApJ, 469, L33
\bibitem{b21} Markwardt C.B., Miller J.M., Wijnands, R., 2002, IAU Circ. 7993
\bibitem{b22} Predehl P., Schmitt J. H. M. M., 1995, A\&A 293, 889    
\bibitem{b23} Remillard R.A. et al., 2002, IAU Circ. 7888
\bibitem{b24} Remillard R.A., Swank, J., Strohmayer T., 2002, IAU Circ. 7893
\bibitem{b25} Rupen M.P., Dhawan, V., Mioduszewski A.J., 2002, IAU Circ. 7893
\bibitem{b26} Strohmayer T.E., Markwardt C.B., Swank J.H., in't Zand J., 2003, 
                ApJ, 596, L67
\bibitem{b27} van Paradijs J.,1983, in Lewan W.H.G., 
                van den Heuvel E.P.J., eds, Accretion-driven Stellar X-ray 
                Sources, Cambridge University Press, Cambridge, p. 191 
\bibitem{b28} van Paradijs J.,  McClintock J.E., 1995, in Lewan W.H.G., 
                van Paradijs J., van den Heuvel E.P.J., eds, X-ray Binaries. 
                Cambridge University Press, Cambridge, p. 58 
\bibitem{b29} Wachter S., Hoard D.W., Bailyn C., Jain R., Kaaret P., 
                Corbel S., Wijnands R., 2000, HEAD, 32, 24.15 
\bibitem{b30} Wang Z. et al., 2001, ApJ, 563, L61
\bibitem{b31} Warner B., Robinson E.L., 1972, MNRAS, 159, 101
\bibitem{b32} Wijnands R., van der Klis M., 1998, Nature, 394, 344
\bibitem{b33} Wijnands R., M\'{e}ndez M., Markwardt C., van der Klis M., 
                Chakrabarty D., Morgan E., 2001, ApJ, 560, 892
\bibitem{b34} Wijnands R., van der Klis M., Homan J., Chakrabarty D., 
                Markwardt C.B., Morgan E.H., 2003, Nature, 424, 44
\end{thebibliography}
\end{document}